# INCREASING DIGITAL INVESTIGATOR AVAILABILITY THROUGH EFFICIENT WORKFLOW MANAGEMENT AND AUTOMATION

Ronald In de Braekt*, Nhien-An Le-Khac†, Jason Farina†, Mark Scanlon†, M-Tahar Kechadi†

*Korps Nationale Politie, Netherlands.
†School of Computer Science, University College Dublin, Ireland.
ronald@braekt.nl, {an.lekhac, mark.scanlon,tahar.kechadi}@ucd.ie,
jason.farina@ucdconnect.ie

*Abstract*

The growth of digital storage capacities and diversity devices has had a significant time impact on digital forensic laboratories in law enforcement. Backlogs have become commonplace and increasingly more time is spent in the acquisition and preparation steps of an investigation as opposed to detailed evidence analysis and reporting. There is generally little room for increasing digital investigation capacity in law enforcement digital forensic units and the allocated budgets for these units are often decreasing. In the context of developing an efficient investigation process, one of the key challenges amounts to how to achieve more with less. This paper proposes a workflow management automation framework for handling common digital forensic tools. The objective is to streamline the digital investigation workflow - enabling more efficient use of limited hardware and software. The proposed automation framework reduces the time digital forensic experts waste conducting time-consuming, though necessary, tasks. The evidence processing time is decreased through server-side automation resulting in 24/7 evidence preparation. The proposed framework increases efficiency of use of forensic software and hardware, reduces the infrastructure costs and license fees, and simplifies the preparation steps for the digital investigator. The proposed approach is evaluated in a real-world scenario to evaluate its robustness and highlight its benefits.

## 1  INTRODUCTION

Information security (Bosworth et al. 2014) concerns protecting information systems against all types of unauthorised access. Despite following best practices, it is almost impossible to eliminate all vulnerabilities from a system. Accepting that fact, digital forensics (A. of Chief Police Officers, Marshall 2008) is about detecting, tracking and identifying the cybercrimes, analysing what has occurred on a system, and providing evidence for successful prosecution. Cybercrime activities today are more difficult to detect due to a range of anti-forensic techniques (Harris 2006). In response to these challenges, an increasing number of digital forensic tools are being released to detect, analyse and report on evidence from a highly specialised medium. For example, Internet Evidence Finder (IEF), created by Magnet Forensics, is a reporting and analysing tool popular with many digital investigators. This tool searches for Internet related artifacts, analyses them and prepares an easily comprehensible standalone report. A detective can then easily find pertinent evidence or indicate the priority of the evidence for further and detailed investigation. There are a number of other popular forensic tools, such as Bulk Extractor (Garfinkel), PhotoRec (Grenier), etc., that are being widely used in the forensic investigating community. On one hand this is an advantage for the investigators as they have more choice in using forensic tools. However, choosing an appropriate tool and exploiting it efficiently for forensic cases are challenges to investigators, as many do not have a Computer Science background. In law enforcement, training is often necessary for using and exploiting new tools.

On the other hand, the growth of hardware storage also poses a significant challenge. For example, the investigation of gigabyte disk images five years ago is now being replaced by terabytes disk images today. The growth of hard drive storage capacity and the diversity in devices greatly increases the time required for the initial acquisition and processing steps of any digital investigation. Investigators find themselves spending more time in acquiring and preparing the evidence instead of investing time in detailed investigation and reporting. The time wasted on acquisition and pre-processing alongside an increasing workload has resulted in long backlogs becoming commonplace. The research problem tackled as part of this paper is to discover how investigators can do more with less cost and effort - helping to improve the efficiency of investigation. Streamlining the digital investigation process and making efficient use of the limited hardware and software can achieve this. In fact, the digital forensic process (Casey 2009, Kent) normally includes four steps: (i) Acquiring/imaging the device; (ii) Preparing the evidence; (iii) Detailed investigation and reporting, and (iv) Clean-up and archiving. Among these steps, the detailed investigation and reporting step is the most important one as this is what ultimately provides evidence for prosecution.

## 1.1 CONTRIBUTION OF THIS WORK

In the context of developing an efficient information system for law enforcement organisations, a workflow management automation framework is proposed as part of this paper for handling digital forensic tools. This framework streamlines the digital investigation process through the execution and control of forensic tools from servers. This optimises the evidence throughput time, as the servers can prepare evidence 24/7. This efficient use of digital forensic software and hardware reduces license fee requirement and simplifies the preparation steps for the digital investigator. The workflow management automation framework presented herein consists of an *Image Creation* component, a *Queue Server Platform* component and a *Clean Up and Archiving* component. The *Image Creation* component simplifies the process of creating a disk image and requires minimal user input for creating the correct folder structure on the file server. When the image is acquired, this component will then create additional jobs in the *Queue Server* resulting in automated evidence pre-processing. The proposed framework is evaluated as part of a case study to determine the robustness of this new approach.

## 2 BACKGROUND AND RELATED WORK

In this section, popular digital forensic tools used in law enforcement organisation are reviewed in the context of the proposed workflow.

### 2.1 Image Acquisition Tools

#### 2.1.1 Encase

To create a hard drive image, investigators commonly use the graphical user interface of EnCase (Guidance Software). EnCase also provides a command line acquisition tool "WinAcq". This tool is designed to run from the command line in the Windows Operating System to acquire whatever physical or logical device you specify. The utility can be run interactively, where it prompts for certain information before it executes the acquisition, or it can be run from the command line with all the options specified on the command line. This allows the tool to be used for scripted operations.

#### 2.1.2 Forensic Toolkit Imager

Another common acquisition tool is Forensic Toolkit (FTK) Imager (AccessData). This tool provides a command line interface and has versions for Linux, Mac and Windows. FTK Images can be run with options specified on the command line - making the tool suitable for scripted operations. This ability to run on multiple operating systems and list the devices makes this tool more suitable for the workflow management framework than the EnCase "WinAcq".

### 2.2 Preparation and Triage Tools

#### 2.2.1 EnCase Portable

EnCase Portable from Guidance Software is a commonly used triage tool. The focus of this tool is for field personnel and not focused on user by a digital investigator. EnCase Portable is a powerful solution, delivered on a USB device, which facilitates forensic professionals and non-experts alike to quickly and easily triage and collect vital data in a forensically sound and court-admissible manner. The remit of this tool is to aid in closing cases faster and reduce the aforementioned backlog by focusing on analysing potential evidence, not searching through data. EnCase Portable is designed to run on not yet confiscated devices.

#### 2.2.2 AD Triage

AD Triage from AccessData, forensically acquires data from both live and powered down computers in the field. With AD Triage, field acquisitions are simplified and do not require a laptop and write blocker. Similar to EnCase Portable, the focus of this tool is for field personnel.

#### 2.2.3 Internet Evidence Finder

Internet Evidence Finder (Magnet Forensics) is a digital forensic software solution used by many forensic professionals to find, analyse and present digital evidence found on computers, smartphones and tablets. This tool is suitable for recovering deleted chat history, social networking communications, webmail, cloud files, browser history, P2P activity, document artifacts, etc. This tool provides the option to run on captured evidence files from the command line.

*2.2.4 Bulk Extractor*

Bulk Extractor (Garfinkel) is a computer forensics tool that scans a disk image, a file, or a directory of files. It extracts useful information without processing the file system or file system structures. The results can be easily inspected, analysed, or processed with automated tools. Bulk Extractor also creates histograms of features that it finds, as features that are more common tend to be more important. Ignoring the file system has the advantage that Bulk Extractor can be used to process any digital media. The program can be used to process hard drives, SSDs, optical media, camera cards, cell phones, network packet dumps, etc. Bulk Extractor is designed to run command line.

*2.2.5 PhotoRec*

PhotoRec (Grenier) is a file data recovery tool designed to recover photos, videos, documents and archives from hard drives and forensic images. It also ignores the file system and is designed to run command line.

## 2.3 Related Work

The abstraction of the digital forensic process was proposed by Reith et al. (2002). However, the authors only described this process as separated steps. Kohn et al. (2008) tried to map the digital forensics process to formal modelling approaches, such as UML, and comment that most of the process models they have reviewed have adopted a more informal approach.

In modelling digital forensic process from a workflow perspective, Wang and Yu (2007) identified the similarities between the software development process and the digital forensic process. However, these two processes are different by their nature. In the digital forensic process, some steps are compulsory and need to be performed in the appropriate order.

At the time of writing, there is no documented framework and very few software tools exist in the literature that are able to control and manage the digital forensic process. In this context, DIALOG (Kahvedzic and Kechadi 2009) is a framework for the management, reuse, and analysis of digital investigation knowledge. DIALOG is a digital investigation ontology that contains the main concepts of digital forensics and their relationships and captures the universe of discourse of the digital investigation domain. It is designed to be independent of any specific investigation and can expand its domain knowledge with definitions of new entities. In fact, this framework is useful to abstract digital investigation knowledge but it cannot be used as a workflow platform.

Wen et al. (2013) proposed a computer forensic workflow management to support the execution of digital forensics on a cloud platform. The objective of this work was to create a Forensics as a Service (FaaS) system. The proposed solution parallelises the creation of disk images by using cloud-computing resources, such as HBase and Hadoop/MapReduce paradigm (Dean and Ghemawat 2008). This approach is not tailored to the sensitivity of transferring potentially incriminating data to a cloud platform. Issues of integrity, consistency and security of evidence could be raised.

# 3 INVESTIGATION ENVIRONMENT AND CURRENT APPROACH

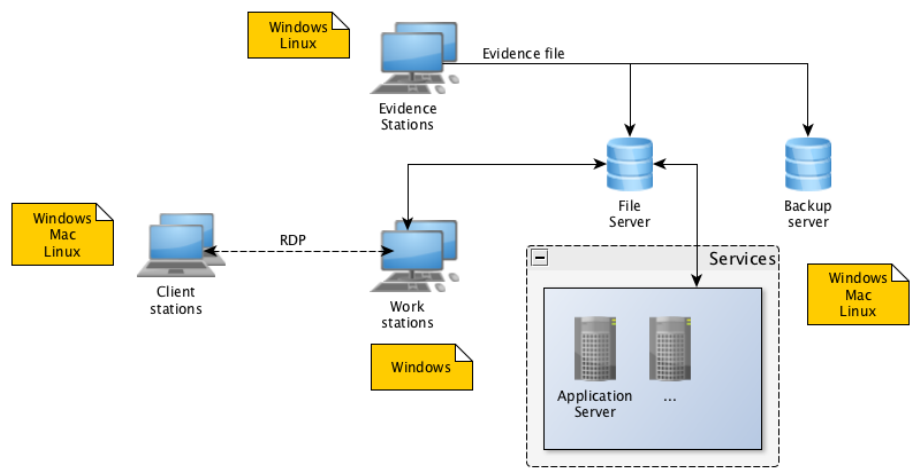

*Figure 1.     Typical Investigation Environment.*

The investigation environment in most law enforcement organisations is built around a file-server that holds the images that are currently being worked on, as can be seen in Figure 1. This typically comprises of a high performance file-server coupled with a backup-server to store copies of the images in case of file-server failure. For disk image creation, there are typically some evidence stations with write blockers for use. For analysing and viewing the evidence there might be several workstations with forensic software. These workstations have two functions: (i) Investigation station for the digital investigator; (ii) Lookout station for the tactical detective. These workstations can be used locally or controlled by a remote desktop system. This makes it possible for the digital investigator to connect to the workstations from other desktops. The tactical detective also uses it locally for report viewing and tactical analysis.

A typical investigation process includes a sequence of steps, as outlined below:
1. Evidence Acquisition - The first step in the investigative process is to create a forensic copy of the data of confiscated devices and to verify the created images. This can generally take several hours or days to complete. The evidence files are stored on the file-server. The verification has to be confirmed before the created image can be copied to the file-server and backup server. Because of the amount of data being shared, this copying process can also take hours to finish. The copy needs to be verified as successful before the locally created image can be deleted from the workstation. Each of these steps have to be checked manually by the digital investigator before the next step can be taken. This is very inefficient and the digital investigator can lose a lot of time during these steps, especially if the image creation or copying is not accomplished during working hours. For this step, streamlining the create image workflow without human intervention will save a significant amount of time and reduce the overall throughput time.
2. Evidence Preparation - Depending on the nature of the crime, a case profile is created with relevant tools. Another time consuming issues experienced here relates to the readiness/availability of these tools. In order to reduce the backlog, it is necessary to involve the tactical detective in an earlier stage of the investigation process.

Automating the preparation of the images would help the tactical investigator to browse pictures, review Internet artifacts and identify who was using the suspect system. This makes it possible to prioritise the devices and cases investigated. A tactical integration of existing digital forensic tools would help reduce costs and help tackle the backlog.

3. Detailed Investigation - In this phase, the digital investigator focuses on analysis and case completion. This analysis is specific for each case and looks for pertinent case related artifacts. It is not possible to automate or streamline this process as due to the diverse range of potential investigations, expert human analysis will always be necessary. However, automatic evidence preparation provided by the framework will support this detailed investigation to be performed faster, as the expert will know where to focus his effort.
4. Reporting - The digital investigator will document all the findings found during the investigation. The faster the detailed investigation is completed, the faster the report can be created and acted upon.
5. Clean-up and Archiving - This is often another time consuming part of the process, because the digital investigator has to manually check all the cases on the file-server against the cases in the registration system. Typically only cases closed for more than 30 days can be cleaned up and made ready for archiving. Case evidence is generally burned to Blu-Ray disc capable of storing 25 GB of data for over 25 years. This is the chosen standard by many law enforcement organisations for the long term archiving of sensitive data.

## 4 PROPOSED WORKFLOW MANAGEMENT AUTOMATION FRAMEWORK

The framework outlined as part of this paper aims to reduce the costs and provide the digital investigator with more time for analysing and completed case reporting. The objectives of the proposed framework are (i) Simplify the investigation process; (2) Decreasing the throughput time; (3) Efficient use of resources and licenses; (4) Quicker results for the tactical detective for browsing images and (5) Browsing Internet history to help identify who has been using the suspect computer.

The framework will simplify the steps needed to be taken and link the investigation process together as much as possible. It will be modifiable, scalable and extendable by the digital investigator and is operating system independent. This workflow management framework is not designed to process all steps mentioned above autonomously. It integrates forensic tools (open source and commercial) for the steps in the investigation process. This workflow management framework is designed to control program execution and verify the resulting output.

The framework consists of three components, which can run independently from each other, but are also attuned to each other: (1) Image creation, for controlling the image creating process; (2) Queue server, for implementing queue servers that will control third party software; and (3) Clean up and archiving component that can run periodically to archive and clean up closed cases.

### 4.1 Image Creation

The goal of the image creation component is to make this process as simple as possible without required any knowledge of the image creation process. Every user that can register a good is able to create an image. This requires minimal user input and implements all the process knowledge and checks into the image creation tool. The user provides information on the confiscated device and the nature of the crime and will be asked to select the preparations steps required. After the acquisition has completed successfully, the preparation jobs will be initiated by the image creation component to streamline the workflow without losing any time between acquiring and preparation.

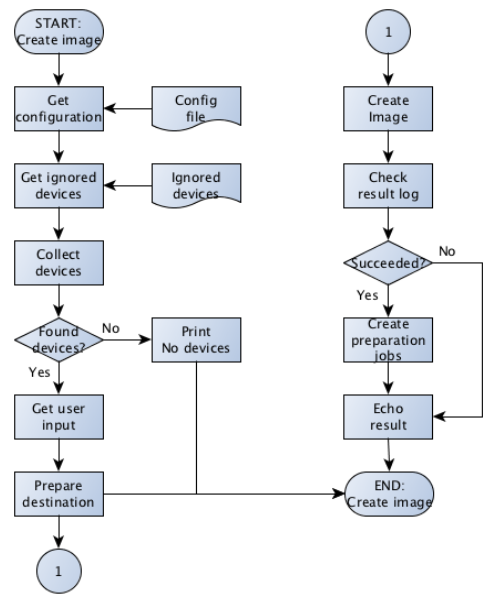

*Figure 2.     Workflow of image creation component.*

Figure 2 shows the global design for the image creation component. This component has a configuration file. The configuration file should be stored on the file server and all imaging machines can use the create image tool with the same configuration file. This makes it possible to easily maintain the available output locations. After the configuration file is read, only newly connected devices are displayed to the user. This prevents the user from inadvertently selecting the incorrect device. When the device is selected, the user will be asked to select the destination, provide a name for the evidence, provide the investigator's credential and the desired preparation. Armed with this information, the script will prepare the output directory, create and check the evidence, and if the created image has succeeded the preparation jobs will be initiated.

### 4.2     Queue Server

The queue server component is designed to monitor a queue folder, and checks on a configurable time interval if there is a job in the queue for processing. A job is a file stored in the queue directory containing specific information about the job. Typically, the job file will specify the location to the source and the location of the stored image file of the confiscated device. The server has a universally compatible implementation centred around a defined folder structure.

The queue will process the jobs in FIFO (First In First Out) order. If the server has found a job file in the queue, it will move the job to the processing folder. After the job is executed the result will be checked. If succeeded, the job file will be moved from the processing folder to the succeeded folder. If failed, the job file will be moved from the processing folder to the failed folder. If the source is locked, the job file will be moved from the processing folder to the locked folder.

Locking the source is implemented to prevent running multiple servers on the same source at the same time. This mechanism is needed because the source evidence is only stored once. This is controlled on the file server for all servers and tools. Multiple tools working on the same source could have a negative influence on the performance of the file server or on processing the image. Writing a predefined lock file to the source location locks the source.

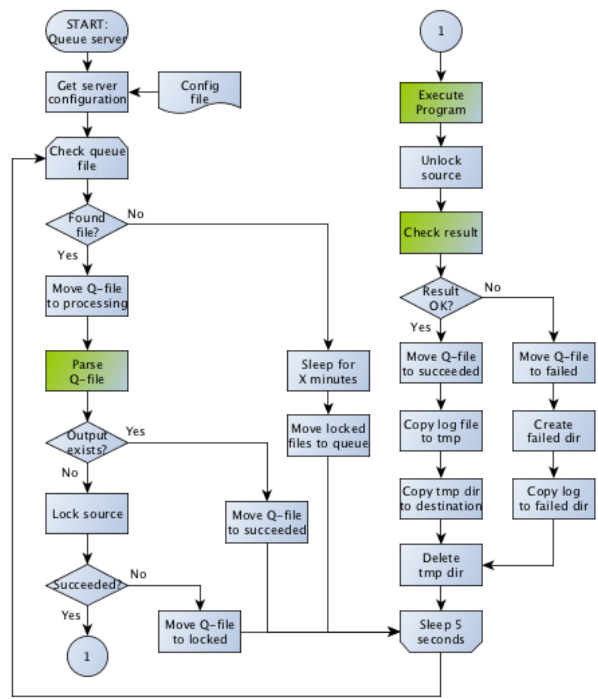

*Figure 3.    Workflow of Queue Server Component*

Figure 3 shows the design of the queue server component. The server is designed to run perpetually. Each server will have his own configuration file and module. This specific module will handle the green collared processes "Parse Q-file", "Execute Program" and "Check result". This module will be dynamically loaded at runtime in the queue server platform. This makes the server logically independent from the type of server, and simplifies the creation of a new server. Briefly, this component has the following advantages:
- Third party software installation, license and updates are only required on the server.
- Execute jobs 24/7.
- Jobs can be initiated from every machine in the network.
- Scalable by duplicating processing servers.
- Processing load is pushed to the server and away from the client machine.

### 4.3 Clean-up and Archiving

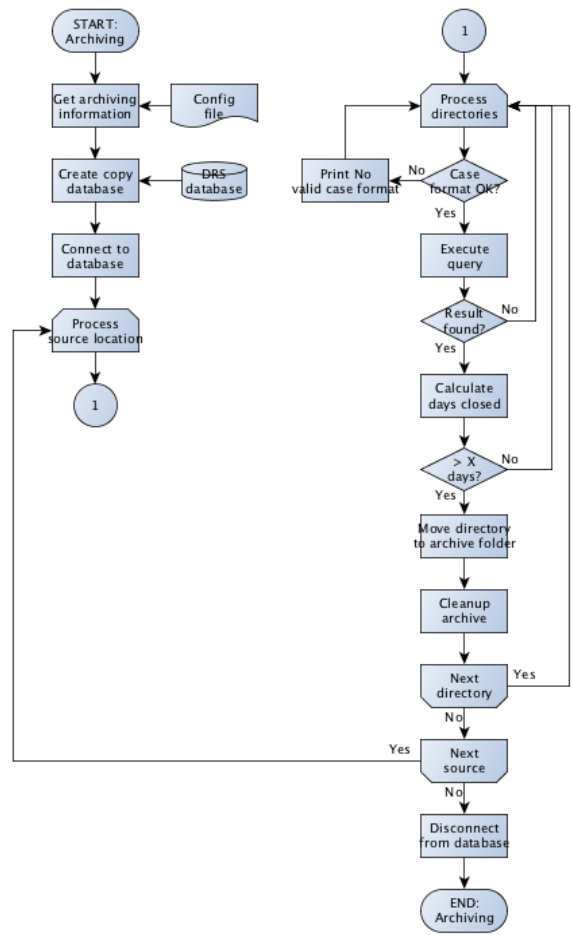

*Figure 4.     Workflow of Clean-up and Archiving.*

The clean-up and archiving component is also designed to run autonomously. The component reads the configuration file, makes a copy of the case registration database DRS and checks each folder in the source directories, as can be seen in Figure 4. If the folder is a case, the case status and date are checked in the database of the case registration system DRS. If the case is "closed" for more than X days the case will be moved to the specified archive folder. The case folder will be cleaned and made ready for archiving.

## 5   FRAMEWORK EVALUATION/CASE STUDY

In this section, an evaluation of the proposed framework is presented through a case study where the framework is implemented in a real-world application in a police department. The case centres around human trafficking (Human trafficking, United Nations 2008). 3 hard drives, 3 laptops and 2 desktops were confiscated as part of the investigation. The tactical detectives were interested in Internet chat communication and the advertised pictures of the victims. This case is used to compare the investigation process using the traditional method and using the new workflow management framework. To calculate the throughput time of the case, the average processing time in GB (gigabyte) per minute for each processing step is used.

### 5.1 Traditional method

Using the traditional method all steps are initiated manually. The digital investigator has to know the internal processes and basic knowledge of the used tooling is required. Preparing the confiscated data for the tactical investigator requires the following steps to prepare the evidence (if two consecutive steps cannot be carried out immediately a "Wait" is inserted):

1) Prepare output directory structure.

2) Creating and verifying the image.

-- Wait --

3) Copy local image to server and backup server.

-- Wait --

4) Delete local copy of the image.

5) Create EnCase case file and start case processor.

-- Wait --

6) The image ready to be viewed.

|  | Evidence system 1 | Evidence system 2 | EnCase case processor |
|---|---|---|---|
| HDD 1 | July 1, 8:00 AM<br>**July 2, 0:40 AM** |  | July 2, 8:00 AM<br>**July 3, 5:20 PM** |
| HDD 2 |  | July 1, 8:00 AM<br>**July 2. 0:40 AM** | July 4, 8:00 AM<br>**July 5, 5:20 PM** |
| HDD 3 | July 2, 8:00 AM<br>**July 4, 10:00 AM** |  | July 7, 8:00 AM<br>**July 11, 12:00 PM** |
| Laptop 1 |  | July 2, 8:00 AM<br>**July 2, 12:10 PM** | July 11, 12:30 PM<br>**July 11, 8:50 PM** |
| Laptop 2 |  | July 2, 12:30 PM<br>**July 2, 5:50 PM** | July 14, 8:00 AM<br>**July 14, 6:40 PM** |
| Laptop 3 |  | July 3, 8:00 AM<br>**July 3, 4:20 PM** | July 15, 8:00 AM<br>**July 16, 0:40 AM** |
| Desktop 1 |  | July 3, 4:20 PM | July 17, 8:00 AM |

|  |  | July 3, 8:30 PM | July 17, 4:20 PM |
|---|---|---|---|
| Desktop 2 |  | July 4, 8:00 AM | July 17, 4:20 PM |
|  |  | July 4, 4:20 PM | July 18, 9:00 AM |

*Table 1.     Throughput time using the traditional method.*

The minimal time an investigator has to wait is the time needed to perform the next step. Additionally, if that step is completed outside of working hours, more time is lost. Table 1 shows an example of a calculation using the traditional method. This example shows the use of two evidence systems and one system to run the EnCase case processor. Each field contains the start and stop time when the next step is finished. If the stop time is outside of work hours (5:00 PM – 8:00 AM) or in the weekend, the next workday at 8:00 AM is chosen to start the next step.

As can be seen in Table 1, the image creation step commences on July 1 at 8:00 AM and the first image can be viewed on July 4 at 8:00 AM. The preparation of the images is completed on July 18 at 9:00 AM. From July 1 at 8:00 AM until July 19 at 9:00 AM, overall it took 409 hours to complete the job.

### 5.2   Using the workflow management framework

Using the workflow management framework, all the preparation is handled automatically after the image is successfully created. The create image tool simplifies the process of making the image. This requires no knowledge about the internal processes or the used tooling. The preparation steps are automatically handled by the servers and need no human interaction. This simplifies the steps we have to make:

1)      Creating the image and selecting the preparation.

-- Wait --

2)      The image ready to be viewed.

The biggest advantage of using the workflow management framework over the traditional way of working is that no time is lost between the steps in the process. The digital investigator only has to focus on creating the images and does not have to remember to start the next step in the process immediately when possible. This saves significant time.

|  | Evidence system 1 | Evidence system 2 | Internet Evidence Finder | Bulk Extractor |
|---|---|---|---|---|
| HDD 1 | July 1, 8:00 AM **July 2, 0:40 AM** |  | July 2, 0:40 AM **July 3, 10:00 AM** | July 2, 0:40 AM **July 2, 4:00 AM** |
| HDD 2 |  | July 1, 8:00 AM **July 2. 0:40 AM** | July 3, 10:00 AM **July 4, 7:20 PM** | July 2, 4:00 AM **July 2, 7:20 AM** |
| HDD 3 | July 2, 8:00 AM **July 4, 10:00** |  | July 4, 7:20 PM **July 8, 11:20** | July 4, 10:00 AM |

|  | AM |  | PM | July 4, 8:00 PM |
|---|---|---|---|---|
| Laptop 1 |  | July 2, 8:00 AM **July 2, 12:10 PM** | July 8, 11:20 PM **July 9, 7:40 AM** | July 2, 12:10 PM **July 2, 1:00 PM** |
| Laptop 2 |  | July 2, 12:30 PM **July 2, 5:50 PM** | July 9, 7:40 AM **July 9, 6:20 PM** | July 2, 5:50 PM **July 2, 6:54 PM** |
| Laptop 3 |  | July 3, 8:00 AM **July 3, 4:20 PM** | July 9, 6:20 PM **July 10, 11:00 AM** | July 3, 4:20 PM **July 3, 6:00 PM** |
| Desktop 1 |  | July 3, 4:20 PM **July 3, 8:30 PM** | July 10, 11:00 AM **July 10, 7:20 PM** | July 3, 8:30 PM **July 3, 9:20 PM** |
| Desktop 2 |  | July 4, 8:00 AM **July 4, 4:20 PM** | July 10, 7:20 PM **July 11, 12:00 PM** | July 4, 8:00 PM **July 4, 9:40 PM** |

*Table 2.        Throughput time using the workflow management system.*

Table 2 shows an example of the same case presented in Section 5.1 using the workflow management system. This example shows the use of two evidence systems and two servers. One server is for running the Internet Evidence Finder and another server is for running the Bulk Extractor. Each field contains the start and stop time when the next step is finished. If the stop time for the image creation is outside of working hours or in the weekend the next workday 8:00 AM is chosen to create the next image.

As can be seen in Table 2, the process is started on July 1 at 8:00 AM. The first results can be viewed on July 2 at 8:00 AM. We finished processing the images on July 11 at 12:00 PM. From July 1 at 8:00 AM until July 12 at 12:00 PM, over all it took 244 hours to complete the job.

5.3 **Interpretation of the Results**

Using the workflow management system, the first results can be viewed after 20 hours versus 57.3 hours with the traditional method. This is 65% quicker with the workflow management system than the traditional way of working.

Calculating the overall throughput time in hours, 409 hours were required using the traditional method and 244 hours using the workflow management system. In this case the workflow management system can save 40% of the overall throughput time with significantly less interaction for the digital investigator - who is now freed to perform other duties during this automation.

## 6 CONCLUSION AND FUTURE WORK

The digital investigation workflow and processes are the same for most organisations, but it is inevitable that the more detailed processes vary from place to place. This is to be expected due to the different investigation environments and working arrangements. The open structure of the proposed platform makes it possible to adapt it to any organisation.

The automation of the preparation process was focused on using existing forensic command line tools that could process images in the EnCase expert witness format. It is not the intention of this framework to replace the detailed, expert-driven investigation process. Earlier analysis of the evidence can result in better prioritisation and increase overall efficiency. With the design of the queue server, any command line tool can be implemented as a job on the queue server.

Overall, the presented workflow management framework has succeeded in streamlining the workflow with the implementation of the queue server. These servers reduce cost by making more efficient use of existing hardware and software. With the automated preparation, cases and images can be prioritised earlier and handled more efficiently. This reduces both the throughput time and helps to tackle the digital evidence backlog. Backup and archiving runs in the background and cleans the case before archiving. Each of these pieces of functionality help eliminate wasted expert investigator time and ultimately provides the digital investigator significantly more time for performing detailed investigation.

### 6.1 Future Work

One limitation of the queue server is that it only supports command line tools. This restricts the queue server from running tools that only have a GUI (Graphical User Interface). Windows focused macro automation tools, such as AutoIt v3 (AutoIt), would make it possible to run any tool as a job in the queue server. The tool will be opened indirectly by the queue server through the execution of a AutoIt v3 script, which handles the simulated keystrokes, mouse movement and window control/manipulation. Popular investigative tools, such as EnCase can be controlled in this manner and can result in saving more preparation time. The implementation of the queue server creates a transparent way of working for the digital investigator.